\documentclass{iopjournal}

\usepackage{graphicx}% Include figure files
\usepackage{dcolumn}% Align table columns on decimal point
\usepackage{bm}% bold math
\usepackage{comment}
\usepackage{xcolor}
\usepackage{bbm}
\usepackage{slashed}
\usepackage[normalem]{ulem}
\usepackage{mathrsfs}
\usepackage{physics}
\usepackage{subcaption}
\usepackage{overpic}

\usepackage{ulem}
\usepackage{soul}

\begin{document}

\articletype{Article type} %	 e.g. Paper, Letter, Topical Review...

\title{Fermionic Signatures of Antispacetime-spacetime Domain Walls}

\author{Justine Charles A. Elveña$^1$\orcid{0000-0000-0000-0000} and Kristian Hauser A.~Villegas$^1$$^*$\orcid{0000-0002-0795-8807}}

\affil{$^1$National Institute of Physics, University of the Philippines Diliman, Philippines.}

% \affil{$^2$National Institute of Physics, University of the Philippines Diliman, Philippines.}

\affil{$^*$Author to whom any correspondence should be addressed.}

\email{kvillegas@nip.upd.edu.ph}


\begin{abstract}
Antispacetime arises naturally in the vielbein formulation of general relativity yet remains invisible to probes coupling only to the metric. We show that domain walls separating spacetime from antispacetime constitute physically distinct and observable structures. Fermions interacting with such walls exhibit striking signatures: scattering induces particle–hole conversion, reflecting tetrad orientation reversal, while the wall hosts localized zero-energy Majorana and chiral modes. In 1+1 dimensions, these modes acquire a pseudoscalar mass term tied to topology and quantum anomalies, providing a direct link to the observability of $\theta$ vacua. These results identify fermionic response as a definitive probe of antispacetime and establish its domain walls as physically detectable.
\end{abstract}

\section{Introduction}
Einstein’s theory of general relativity may be formulated alternatively in terms of the vielbein rather than the metric tensor \cite{Wald1984}. In this formulation, the gravitational action is proportional to the determinant of the vielbein, $\det(e)$, which allows for contributions of either sign. By contrast, the Einstein–Hilbert action depends only on the positive square root of the metric determinant, which is equal to $\abs{\det(e)}$ \cite{Rovelli2012}. Moreover, the vielbein $e^{\hat{a}}_{\mu}(x)$ provides a local mapping between spacetime coordinates and orthonormal inertial frames, enabling a consistent definition of the Dirac equation in curved spacetime \cite{Wald1984}. The necessity of the vielbein formulation for coupling gravity to fermions, which are ubiquitous in nature, supports the view that it represents a more fundamental framework than a purely metric description.

One striking consequence of the vielbein formulation is the emergence of antispacetimes, corresponding to regions in which $\det(e)<0$. Contributions from such regions to the gravitational path integral are analogous to backward-propagating amplitudes in quantum theory, suggesting an interpretation of antispacetimes as the spacetime counterparts of antiparticles \cite{Rovelli2012}.

The existence of antispacetimes carries profound and far-reaching implications such as the reversibility of causality \cite{Donoghue2019} and the existence of negative local temperature \cite{Volovik2019NegativeTemperature}. In cosmology, extensions of the standard Big Bang model have proposed the creation of a universe–antiuniverse (spacetime-antispacetime) pair, offering a potential explanation for unresolved problems such as the baryon asymmetry and the nature of dark matter \cite{Boyle2018, Kumar2025}.

Whether antispacetimes exist in nature is still an open question. Adding more challenge is the fact that they can not be detected using conventional probes that couple only to the metric. This points to the importance of fermions to probe antispacetimes. In \cite{Christodoulou2012_antispacetime}, it was shown that a localized region of antispacetime could, in principle, be detected through interference between Dirac fermions of different masses. Here, we propose a more direct way to probe an antispacetime and show that the interplay between fermions and antispacetime-spacetime domain wall (ASDW) exhibits rich phenomenology. In particular, we demonstrate that scattering across the wall exhibits an anomalous particle–hole conversion and that the Dirac equation admits localized solutions at the ASDW. Notably, we show that Majorana zero modes, chiral fermions, and pseudoscalar mass can emerge among these localized states. These provide clear and potentially observable signatures of the ASDW.

\section{The Dirac equation with ASDW}
We are interested in the solutions of the Dirac equation,
\begin{align}
(i\gamma^{\hat{\alpha}}e^\mu_{\hat{\alpha}}\mathcal{D}_\mu-m)\psi=0,  
\end{align}
in a background containing a domain wall
separating regions of spacetime and antispacetime. Here, $\gamma^{\hat{\alpha}}$ are the gamma matrices obeying the Clifford algebra, $e^\mu_{\hat{\alpha}}$ is the inverse vielbein with coordinate index $\mu$ and Lorentz index $\hat{\alpha}$, and $\mathcal{D}_\mu=\partial_{\mu}+\Gamma_\mu$ is the covariant derivative for spinors. The spinor affine connection is given by
\begin{align}
\Gamma_\mu=\frac{1}{8}\omega_{\mu\hat{a}\hat{b}}[\gamma^{\hat{a}},\gamma^{\hat{b}}],
    \end{align}
where 
$\omega_{\mu \hat{c}}^{\hat{a}} = \Gamma _{\mu\alpha}^\nu e^{\hat{a}}_\nu e^\alpha_{\hat{c}}-e^\nu_{\hat{c}}\partial_\mu e_\nu ^{\hat{a}}$ is the spin connection coefficient, with $\Gamma _{\mu\alpha}^\nu$ being the Christoffel symbols.

Let us first consider the simplest case in 1+1 dimensions, where the vielbeins are 
\begin{align}
\label{eq:asdw1+1}
    e_0^{\hat{0}}=1,\;\;\;\;e_1^{\hat{1}}=\tanh(ax).
\end{align}
The latter switches sign across the ASDW at $x=0$, so that for $a > 0$ ($a < 0$) there is antispacetime (spacetime) in the region $x<0$ and spacetime (antispacetime) in the region $x>0$. The parameter $a^{-1}$ sets the length scale of the transition between the two regions, which we speculate to be at the quantum gravity scale. Figure \ref{fig:plots} (a) shows the profile of the wall. The background corresponds to the metric
\begin{align}
    ds^2 = dt^2 - \tanh^2(ax)dx^2,
\end{align}
which is Minkowski, as can be seen with the coordinate transformation $\dd{\xi} = \tanh(ax)\dd{x}.$ Notice that this metric is degenerate. Such metrics are accommodated in Ashtekar's polynomial formulation of gravity, which relies on the connection and triad variables, and the absence of a defined inverse metric is permissible \cite{Bengtsson}.

The proper distance from any nonzero point $x$ to the ASDW is finite, and the null geodesics satisfy 
\begin{align}
    t=\pm a^{-1}\ln\cosh(ax)+C, 
\end{align}
with real constants $C$. These imply that the ASDW is accessible by observers from either spacetime or antispacetime, and communication across the domain wall is possible. This differs from the inaccessible domain walls in analogue ASDW systems realized in superfluid helium. We briefly discuss these analogue ASDW systems in Sec. \ref{sec:ASDW Analogues}.

Note also that, in contrast to the well-known Jackiw–Rebbi topological defect \cite{Jackiw1976}, where the domain wall arises from the change in sign of the Dirac mass term $m(x)$, the domain wall in our set-up emerges from the structure of the kinetic term of the Dirac equation, $i\gamma^{\hat{\alpha}}e^\mu_{\hat{\alpha}}\mathcal{D}_\mu\psi$, through the spatial dependence of the vielbein. Nonetheless, as demonstrated in studies of analogue antispacetimes in superfluid $^3$He, such a structure can constitute a topological wall \cite{Silveri2014HardDomainWalls, Salomaa1988CosmiclikeDomainWalls}.

The only nonzero Christoffel symbol in this background is $\Gamma_{xx}^x = a \sech(ax) \csch(ax),$ and the spin connection coefficients all vanish, which simplifies the covariant derivative $\mathcal{D}_\mu \to \partial_\mu$. With the following representation for gamma matrices, $ \gamma^{\hat{0}}=\sigma^z,\gamma^{\hat{1}}=-i\sigma^y,$ the Dirac equation reads
\begin{align}\label{eq:1+1Diraceq}
    \big(i\sigma^z\partial_t + \coth(ax)\sigma^y\partial_x - m\big)\psi = 0.
\end{align}
We take solutions of the form
\begin{align}
    \psi(t,x) =  \tilde{\psi}_{+}e^{-i\omega t} + \tilde{\psi}_{-}e^{i\omega t},
\end{align}
where $\omega$ is taken to be greater than zero, so that $\tilde{\psi}_{+}$ and $\tilde{\psi}_{-}$ are the positive and negative energy modes, respectively. Writing their components as $\tilde{\psi}_{\pm}=(\tilde{\psi}_{A\pm},\tilde{\psi}_{B\pm})^T$, we obtain the following coupled differential equations
\begin{equation}\label{diraccomponents}
	\begin{aligned}
		(\pm\omega-m)\tilde{\psi}_{A\pm}-i\coth (ax) \partial_x \tilde{\psi}_{B\pm}=0\\
		(\mp\omega-m)\tilde{\psi}_{B\pm} + i\coth (ax) \partial_x \tilde{\psi}_{A\pm}=0.
	\end{aligned}
\end{equation}
This can be solved for $\tilde{\psi}_{A\pm}$ and $\tilde{\psi}_{B\pm}$ to get the overall solution
\begin{align}
\label{eq:gensoln}
    \psi = a_\lambda e^{-i\omega t} \cosh^{\lambda}(ax)u_{\lambda} + b^*_\lambda e^{i\omega t} \cosh^{-\lambda}(ax)v_{\lambda},
\end{align}
where $(a\lambda)^2=m^2-\omega^2.$ Here, $a_\lambda$ and $b^*_\lambda$ are the integration constants, which can be promoted into particle annihilation and antiparticle creation operators, respectively, via second quantization. The spinors $u_\lambda$ and $v_\lambda$ are
\begin{align}
\label{eq:spinors}
    u_\lambda = N\qty(1, \frac{i\lambda a}{m+\omega})^T,\;\;\;
    v_\lambda= N\qty(\frac{i\lambda a}{m+\omega}, 1)^T,
\end{align}
where $N$ is a normalization constant (See Appendix \ref{sec:1+1detailedSoln}).

From this general solution, we identify two classes of solutions with markedly different properties. The first consists of delocalized states that extend to $x \rightarrow \pm\infty$, where they asymptotically reduce to the familiar plane-wave solutions of the free Dirac equation \cite{Guilarte2019}. The second consists of solutions localized at the ASDW. Let us analyze each class of solutions in turn.

\section{Scattering Solution} 
We will now show that imposing the continuity condition to the solutions of the Dirac equation implies the particle-hole conversion during scattering at the ASDW. Let us consider the asymptotic form of Eq. \eqref{eq:gensoln} and focus on the $a > 0$ case and positive energy solutions to keep the discussion simple and clear. The result for negative energy solutions, outlined in Appendix \ref{sec:negativeEnergyscattering}, is similar. Far from the ASDW, the solutions must obey the plane wave dispersion relation $\omega^2=k^2+m^2$. We therefore have $\omega>m$, so that 
\begin{align}\label{eq:lam1+1}
    \lambda = \pm a^{-1}\sqrt{m^2-\omega^2}= \pm ika^{-1}.
\end{align}
In the limit $x\to -\infty$, Eq. \eqref{eq:gensoln} then becomes
\begin{align}\label{eq:-infGenplusE}
    \psi=\frac{a_\lambda e^{-i\omega t-ikx}}{2^{ik/a}}
    u_{ik/a} + \frac{a_{-\lambda} e^{-i\omega t+ikx}}{2^{-ik/a}}u_{-ik/a}.
\end{align}
Here, we take $k$ to be strictly nonnegative and put explicit signs $\pm k$, to distinguish between left- and right-moving solutions. 

Without loss of generality, we consider the case where there is an incident right-moving wave from the antispacetime region 
\begin{align}
\label{eq:leftsolution}
\psi=&e^{-i\omega t+ikx}u_{-ik/a}+Re^{-i\omega t-ikx}u_{ik/a}.
\end{align}
The first term above is the incident state with a unit amplitude. Meanwhile, the second term corresponds to the reflected state, and $R$ is the reflection amplitude. 

In order for Eq. \eqref{eq:-infGenplusE} to have this form, we must have
\begin{align}
\label{eq:stitch1cond}
    a_{-\lambda} = 2^{-ik/a}\;\;\text{and}\;\; a_\lambda 2^{-ik/a} = R.
\end{align}

Next, we consider the region $x\rightarrow+\infty$ where Eq. \eqref{eq:gensoln} reduces to
\begin{align}
\label{eq:+infGenplusE}
    \psi=\frac{a_\lambda e^{-i\omega t+ikx}}{2^{ik/a}}u_{ik/a} +  \frac{a_{-\lambda} e^{-i\omega t-ikx}}{2^{-ik/a}}u_{-ik/a}.
\end{align}

If we assume that there is only a right-moving transmitted wave so that Eq.~\eqref{eq:+infGenplusE} takes the form 
\begin{align}
    \psi = T_R e^{-i\omega t+ikx}u_{ik/a},
\end{align}
where $T_R$ is the transmission amplitude, then we have $a_{-\lambda} = 0$. This contradicts the first of Eqs.~\eqref{eq:stitch1cond}, since by assumption the incident amplitude is nonvanishing. Hence, the assumption that only a right-moving plane-wave is transmitted to $x\rightarrow +\infty$ is inconsistent.

If we generalize the transmitted plane-wave to contain both right-moving and left-moving parts, then the constants in Eq. \eqref{eq:+infGenplusE} should be 
\begin{align}
   a_\lambda2^{-ik/a}=T_R\;\;\text{and}\;\;a_{-\lambda}2^{ik/a} = T_L,  
\end{align}
where $T_R$ and $T_L$ are the transmission amplitudes for the right and left-moving waves, respectively. This is now consistent with the presence of an incident wave. Comparing this to Eq. \eqref{eq:stitch1cond} gives a relation on the amplitudes of left and right wave transmission $T_R = R$ and $T_L = 1$. 

To gain further insights on the transmission and reflection probabilities, we now calculate the currents of the asymptotic incident, reflected, and transmitted states along the scattering direction $j^x = \bar{\psi}\gamma^x\psi.$ Taking note that $\gamma^x$ is a curved gamma matrix containg a $\coth(ax)$ factor, which tends to $\pm 1$ at $x\rightarrow\pm\infty,$ we get the following
\begin{equation}
\begin{aligned}
    &j_{\text{inc}}^x = 2k \\
    &j_{\text{ref}}^x =-2kRR^* \\
    &j_{\text{trans,r}}^x = 2k T_R T_R^* = 2kRR^*\\
    &j_{\text{trans,l}}^x = -2k T_LT_L^* = -2k.
\end{aligned}
\end{equation}

From these currents, the probabilities to be reflected ($P_R$), transmitted as right-moving wave ($P_{T,\text{right}}$), and transmitted as left-moving wave ($P_{T,\text{left}}$) are, respectively
\begin{equation}
\begin{aligned}
    &P_R = \frac{|j_{\text{ref}}|}{j_{\text{inc}}} = |R|^2\\
    &P_{T,\text{right}} =\frac{|j_{\text{trans,r}}|}{j_{\text{inc}}} =|R|^2\\
    &P_{T,\text{left}} = \frac{|j_{\text{trans,l}}|}{j_{\text{inc}}} = 1.
\end{aligned}
\end{equation}
Observe that the sum of the quantities above is $1+2|R|^2 > 1$, which means that for $|R|^2>0$, the total probability is not properly normalized to unity, and $P_R$, $P_{T,\mathrm{left}}$, and $P_{T,\mathrm{right}}$ can not be interpreted as probabilities. Although this may suggest pair creation by analogy with the Klein paradox \cite{Calogeracos}, such an interpretation is not feasible. Unlike the Klein setup, there is no tunable step potential, but rather an antispacetime–spacetime domain wall. Moreover, in the Klein paradox, the transmission probability becomes negative, enabling an antiparticle interpretation, whereas here $P_R = P_{T,\mathrm{right}} > 0$.

We therefore impose $R = T_R = 0$, for which the total probability is normalized. Since $P_{T,\mathrm{left}} = 1$, transmission across the domain wall is perfect, albeit into a wave propagating with the opposite momentum.

This result runs counter to standard scattering intuition. Physically, it originates from a reversal of spatial orientation across the interface. Mathematically, this is reflected in the asymptotic behavior of the $\cosh^{ik}(ax)$ factors appearing in the general solution Eq. \eqref{eq:gensoln}. In the limits $x \to \pm\infty$, these terms reduce to plane waves $e^{\pm ikx}$ with opposite signs, such that a right-moving mode in the antispacetime region is transmitted as a left-moving mode in the spacetime region, and vice versa.

Our result can further be interpreted within the Feynman–Stückelberg framework \cite{Feynman1949}, in which a right-moving particle may equivalently be viewed as a left-moving hole. In this picture, an incident particle is transmitted across the domain wall as an antiparticle and vise versa. This process is reminiscent of Andreev reflection at a normal-metal–superconductor interface, where an incident electron is converted into a hole \cite{Andreev1964}. A similar analogy appears in quantum descriptions of Hawking radiation at a black hole horizon \cite{ManikandanBlackholesAndreev}. In the present case, however, particle–hole conversion occurs upon transmission rather than reflection, and perfect transmission replaces the perfect reflection characteristic of subgap Andreev processes. These results hold in higher dimensions of scattering processes occurring perpendicular to the wall.

\subsection{Regularity of the solution at the ASDW}
The continuity of the solution at the ASDW (located at $x=0$ in our configuration) is essential in our demonstration of the particle–hole conversion during scattering across the wall. We show that the solution is regular and necessarily continuous at the ASDW, despite the Dirac equation containing a Dirac-delta-like term, $\coth(ax)\partial_x$, originating from the vielbein. The easiest way to show this is by replacing $\psi(x,t)\rightarrow e^{-i\omega t}\chi(x)$ then multiplying by $\tanh(ax)$ to convert the Dirac equation into
\begin{align}
i\gamma^1\partial_x\chi+\tanh(ax)(\omega\gamma^0-m)\chi=0,
\end{align}
which is clearly regular at $x=0$. Alternatively also, we can decouple Eq. \eqref{diraccomponents} to obtain a second-order ordinary differential equation, which has the form 
\begin{align}
    \frac{d^2\chi_i}{dx^2}+P(x)\frac{d\chi_i}{dx}+Q(x)\chi_i=0,
\end{align}
where
\begin{align}
P(x)=-a\sech^2(ax)\coth(ax),\;\;\; Q(x)=(\omega^2-m^2)\tanh^2(ax).
\end{align}

The following limits
\begin{align}
    \lim_{x\rightarrow 0}xP(x)=\frac{1}{a},\;\;\;
     \lim_{x\rightarrow 0}x^2Q(x)=0
\end{align}
are finite, showing that $x=0$ is a regular singular point \cite{ArfkenWeber2005}. Since the general scattering solution remains finite and nonvanishing at $x=0$, it follows that it is continuous at that point. This demonstrates that the presence of the ASDW is fundamentally different from the case of the Dirac equation with Dirac-delta potentials.

\section{Localized solutions} 
The other class of solutions from Eq. \eqref{eq:gensoln} have $\omega<m$. These solutions do not reduce to plane wave solutions far from the domain wall, but are instead localized. For convenience, we redefine $\lambda$ to take only the positive root in the first equality of Eq. \eqref{eq:lam1+1}.We add a minus sign, $-\lambda$, to denote what was previously its negative value. 

\begin{figure}[htb]
	\centering
    \begin{overpic}[width=0.5\textwidth]{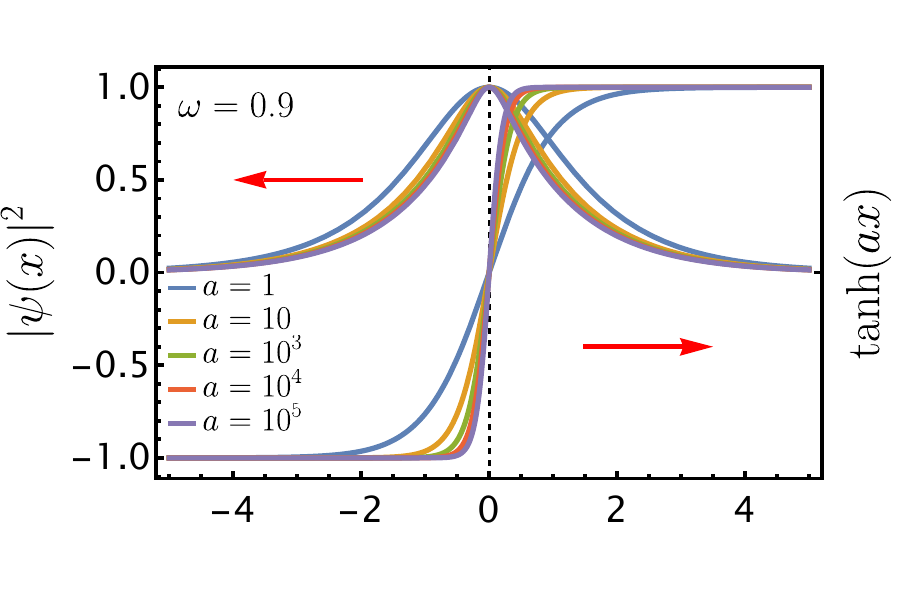}
		\put(0,60){\textbf{(a)}} 
        \put(53,5){\textbf{x}}
	\end{overpic} \hfil
    \raisebox{0.3cm}{
	\begin{overpic}[width=0.45\textwidth]{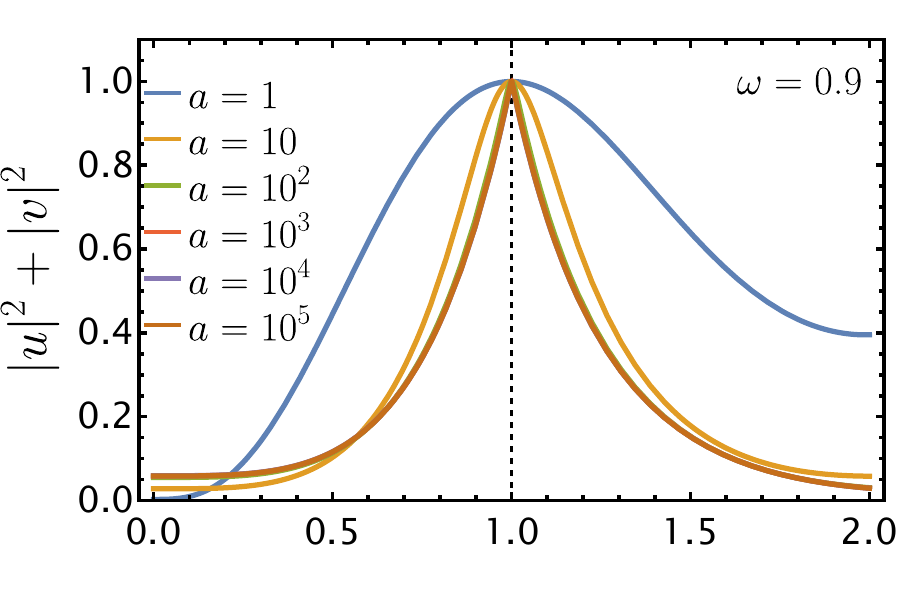}
		\put(0,60){\textbf{(b)}} 
        \put(57,2){\textbf{r}}
	\end{overpic}}% 
	\caption{Probability density illustrating localization near an ASDW. (a) Positive-energy eigenstate localized at the planar ASDW at $x=0$, together with the vielbein profile $e_1^{\hat{1}}=\tanh(ax)$. (b) Probability density for a fermion with $\omega=0.9$ near a spherical ASDW with radius $r_0=1$, shown for several values of $a$. In both panels we set $m=1$.}
	\label{fig:plots}%
\end{figure}

In the case of $a > 0,$ we retain only the spinors in Eq. \eqref{eq:gensoln} with the factor $\cosh^{-\lambda}(ax)$. These terms decay exponentially $\sim e^{-\lambda|x|}$ at $x\to\pm\infty$. Explicitly, the bound state is 
\begin{align}\label{eq:bound1+1}
    \psi= \cosh^{-\lambda}(ax)\left[ a_{-\lambda} e^{-i\omega t}u_{-\lambda}(\omega) + b^*_\lambda e^{i\omega t}v_\lambda(\omega) \right].
\end{align}
Figure \ref{fig:plots} (a) shows the probability densities for various $a$, indicating increased localization of the fermions with increasing $a$.

At zero energy, the spinors in Eq.~\eqref{eq:spinors} reduce to 
\begin{align}
  u_{-\lambda}(0)=N(1,-i)^T\;\;\text{and}\;\;v_{\lambda}(0)=N(i,1)^T.   
\end{align}
Up to an irrelevant overall phase, these zero-energy solutions are invariant the under charge conjugation operator $CK,$ where $C = \sigma^x$ and $K$ is complex conjugation, 
\begin{align}
    Cu^*_{-\lambda}(0)=iu_{-\lambda}(0)\;\;\text{and}\;\;Cv^*_{\lambda}(0)=iv_{\lambda}(0). 
\end{align}
These are therefore Majorana zero modes, analogous to the boundary states arising at topological interfaces \cite{HasanKane2010, QiZhang2011}.

On the other hand, for $a < 0,$ the bound solution is made up of terms from Eq. \eqref{eq:gensoln} with a factor of $\cosh^{\lambda}(ax):$
\begin{align}
    \psi= \cosh^{\lambda}(ax)\left[ a_{\lambda} e^{-i\omega t}u_{\lambda}(\omega) + b^*_{-\lambda} e^{i\omega t}v_{-\lambda}(\omega) \right].
\end{align}
These are similarly Majorana modes.

\subsection{Higher dimensional ASDWs}
We now extend the analysis to 1+1 and 2+1-dimensional ASDWs. Aside from Majorana zero modes, we find that the ASDW hosts chiral fermions unidirectionally propagating at the speed of light. Moreover, unlike in 0+1 ASDW where there is no spatial dimension, the localized modes in higher-dimensional ASDWs carry additional spatial degrees of freedom, allowing us to construct an effective boundary Lagrangian that captures the dynamics of these localized modes. In the following discussion, we focus primarily on the results for $a > 0$ and only briefly address the $a < 0$ case. Unless otherwise specified, the results for $a < 0$ are identical to those for $a > 0.$

Let us consider the case of a 1+1D ASDW first. We place the ASDW along the x axis and use the vielbeins $ e^{\hat{0}}_0 = 1$, $e^{\hat{1}}_1 = 1$, and $e^{\hat{2}}_2 = \tanh(ay)$. Again, the background is Minkowski spacetime, and the spin connection vanishes. Using the following representation for flat gamma matrices $\gamma^{\hat{0}} = \sigma^3,\gamma^{\hat{1}} = -i\sigma^1, \gamma^{\hat{2}} = -i\sigma^2,$
the Dirac equation reads
\begin{align}\label{eq:2+1dirac}
	(i\sigma^3 \partial_t + \sigma^1 \partial_x +\sigma^2 \coth(ay) \partial_y - m) \psi = 0.
\end{align}

The localized solution is given by
\begin{align}\label{eq:bound2+1}
    \psi=\int\frac{\dd{k_1}}{\sqrt{2\pi}}N\cosh^{-\lambda}(ay)\left[a_{-\lambda} e^{-i\omega t+ik_1x} u_{-\lambda}(\omega) +  b^*_\lambda e^{i\omega t - ik_1 x}v_\lambda(\omega) \right],
\end{align}
where we generalized $\lambda =  a^{-1}\sqrt{m^2+k_jk^j - \omega^2}$ to higher dimensions. The Einstein summation in $k_jk^j$ runs over the momentum components parallel to the domain wall. The spinors are still given by Eq. \eqref{eq:spinors}, but with $\lambda a$ replaced by $\lambda a + k_1$ and $\lambda a-k_1$ for $u_{\lambda}$ and $v_{\lambda},$ respectively.

Recall that the Dirac Lagrangian in curved spacetime is given by
\begin{align}\label{eq:origLagrangian}
    \mathcal{L}=\sqrt{-g}\bar{\psi}\;(i\gamma^{\hat{a}}e^{\mu}_{\hat{a}}\mathcal{D}_{\mu}-m)\psi.
\end{align}
where $\sqrt{-g}=|\det(e)|=|\tanh(ay)|$ is the positive root of the metric determinant and $\bar{\psi}=\psi^\dagger\gamma^{\hat{0}}$ is the Dirac adjoint of $\psi$.
To obtain the effective Lagrangian for the bound states, we separate the factor that localizes the modes at the domain wall and write the solution as 
\begin{align}
    \psi(t,x,y) = \cosh^{-\lambda}(ay)\chi(t,x), 
\end{align}
where $\chi(t,x)$ is a two-component spinor. Plugging this into Eq. \eqref{eq:origLagrangian} and factoring out the $y$ integral from the action, we obtain the ASDW Lagrangian
\begin{align}\label{eq:1+1bdyL}
    \mathcal{L}_{1+1\;DW} = \bar{\chi}(i\gamma^{\hat{j}}\partial_j-  \lambda a\sigma^2-m)\chi.
\end{align}

Because of the even dimensionality of the domain wall, it has a chiral operator $\gamma^5\equiv\gamma^{\hat{0}}\gamma^{\hat{1}}=\sigma^2$ \cite{KaplanChiralityLecture}. The localized fermions are therefore described by an effective Lagrangian with a pseudoscalar mass term $m_5 \bar{\chi}\gamma^5\chi$, with $m_5 = \lambda a$. The absence of the usual factor of $i$ in the pseudoscalar mass term suggests that the boundary Lagrangian is non-Hermitian. However, it is in fact a valid Lagrangian because it is Hermitian up to a total derivative 
\begin{align}
	\mathcal{L}^{\dagger}_{2+1\;DW} &=|\tanh(ay)|\cosh^{-2\lambda}(ay)[
	-i\partial_t\chi^{\dagger}\sigma^3\sigma^3 \chi + \partial_x\chi^\dagger \sigma^1\sigma^3\chi-\lambda a\chi^{\dagger}\sigma^2\sigma^3 \chi - m\chi^{\dagger}\sigma^3\chi]\\
    &= |\tanh(ay)|\cosh^{-2\lambda}(ay)\bar{\chi} \quad(i\sigma^3\partial_t + \sigma^1 \partial_x +  \lambda a\sigma^2-m)\chi  \\
	&= |\tanh(ay)|\cosh^{-2\lambda}(ay)\bar{\chi} (i\sigma^3\partial_t + \sigma^1 \partial_x -  \lambda a\sigma^2-m)\chi  + \partial_y[\cosh^{-2\lambda}(ay)]\chi^\dagger\sigma^2\sigma^3\chi \\
	&= \mathcal{L}_{2+1\;DW}  \; + \; \text{full derivative term},
\end{align}
where in going to the second line, we integrated by parts, and in going to the last line, we identified $\partial_y[\cosh^{-2\lambda}(ay)]\chi^\dagger\sigma^2\sigma^3\chi$ as a total derivative term.

The presence of a pseudoscalar mass term is particularly noteworthy. When coupled to gauge interactions, it leads to a rich interplay among discrete symmetry breaking, quantum anomalies, and topology. In quantum chromodynamics, for instance, such a term is intimately connected to strong CP violation and to the physical observability of the $\theta$ vacuum \cite{Jackiw1976vacuum, Callan1976, Peccei1977, Witten1979, Veneziano1979}. The possibility that superfluid $^3$He-A can host analogues of ASDW, together with emergent pseudoscalar mass and gauge symmetry \cite{Volovik2009}, provides a potential platform for investigating these fundamental phenomena in controlled, table-top experiments.

Remarkably, there are fermionic modes that are effectively massless  in the ASDW even though they are massive in the bulk. This happens when the scalar and pseudoscalar masses cancel each other. To see this, observe that from the domain wall Lagrangian Eq. \eqref{eq:1+1bdyL}, solutions could satisfy a Weyl equation
\begin{align}\label{eq:1+1Weyl}
    (i\sigma^3\partial_t+\sigma^1\partial_x)\chi = 0
\end{align}
if they also follow the relation
\begin{align}
    \sigma^2\chi=-\frac{m}{\lambda a}\chi.
\end{align}
Note that $\sigma^2$ has eigenvalues $\pm1,$ and the equation above implies $\lambda a = m$
and that $\chi$ is an eigenstate of $\sigma^2$ with eigenvalue $-1.$ Consequently, with $\sigma^2$ being the chiral operator in the ASDW, these fermions are left-chiral. If we plug in the definition of  $\lambda=a^{-1}\sqrt{m^2 + k_jk^j-\omega^2}$ into the relation $\lambda a=m$, we also confirm that these modes have a massless dispersion $\omega^2=k^2_1,$ and thus propagate at the speed of light. 

Additionally, the solutions are left-moving. This is seen by left-multiplying Eq. \eqref{eq:1+1Weyl} by $\sigma^3$ and replacing $\sigma^2$ with its eigenvalue, yielding
\begin{align}
(\partial_t-\partial_x)\chi = 0,
\end{align}
with solution $\chi = f(x+t).$

Had the case been $a<0,$ the bound solutions would have a factor of $\cosh^{\lambda}(ay),$ and the effective Weyl modes would need to satisfy $\sigma^2\chi=m/(\lambda a)\chi.$ This corresponds to right-chiral modes that will be right-moving along the ASDW.

This result shows that the localized fermionic modes not only signal the presence of the ASDW, but also determine which side corresponds to spacetime and which to antispacetime, through their chirality and propagation direction. We also note that, for charged fermions, having such unidirectionally propagating chiral modes will create a supercurrent in the ASDW in the presence of an electric field along the ASDW, similar to axion strings. This introduces another signature for ASDWs because this has observable effects when passing through astrophysical and
primordial magnetic fields \cite{superconductingStrings, Bagherian2024, Fukuda, Ibe}.

Finally, the static zero mode solutions $\omega = k_1 = 0$ are Majorana fermions. Using the  solution Eq. \eqref{eq:bound2+1},
\begin{align}
	u_{-\lambda}(\omega=0,k_1=0) &=(1, \; -i)^T\\
	v_{\lambda}(\omega=0,k_1=0) &= (i, \; 1)^T,
\end{align}
and the charge conjugation operator given by $CK,$ where $C =i\sigma^1,$ we see that the spinors follow the Majorana condition up to a phase
\begin{align}
	Cu_{-\lambda}^*(\omega=0,k_1=0) &=(-1, \; i)^T = -u_{-\lambda}(0,0)\\
	Cv^*_{\lambda}(\omega=0,k_1=0) &= (i, \; 1)^T = v_{-\lambda}(0,0).
\end{align}

We now consider the more realistic case of $2+1$-dimensional ASDW in a 3+1D universe. We take the vielbeins to be
$e^{\hat{0}}_0=e^{\hat{1}}_1=e^{\hat{2}}_2=1,$ and $e^{\hat{3}}_3=\tanh (az),$ such that the ASDW is placed in the x-y plane. Using the chiral representation for the gamma matrices,
\begin{align}\label{eq:chiralRep}
	\gamma^\mu=
	\begin{pmatrix}
		0 & \sigma^\mu\\
		\bar{\sigma}^\mu & 0
	\end{pmatrix}
	,\;\; \sigma^\mu=(1,\vec{\sigma}),\;\;\bar{\sigma}^\mu=(1,-\vec{\sigma}),
\end{align}
where the vector components of $\vec{\sigma}$ are the Pauli matrices.
The localized solution has the same form as Eq. \eqref{eq:bound2+1}, with replacements $\cosh^{-\lambda}(ay)\rightarrow\cosh^{-\lambda}(az)$, $dk_1\rightarrow dk_xdk_y$, and the extension of $k_jk^j$ to two dimensions. There is also an additional sum over the spins $\sum_{\chi=\pm}$. The spinors are now given by 
\begin{equation}
    \label{eq:3+1genSolns}
    \begin{aligned}
    u_{-\lambda}(\omega,\mathbf{k})=&(\chi,m^{-1}(\omega+k_j\sigma^j+i\lambda a\sigma^3)\chi)^T\\
    v_\lambda(\omega,\mathbf{k})=&(m^{-1}(-\omega+k_j\sigma^j-i\lambda a\sigma^3)\chi,\chi)^T. 
\end{aligned}
\end{equation}

We can again factor the $z$-dependence of the solution 
\begin{align}
    \psi(t,\mathbf{r})=\cosh^{-\lambda}(az)\phi(t,x,y),
\end{align}
where $\phi$ is a four-component spinor, to get the effective Lagrangian in 2+1-dimensional wall
\begin{align}\label{eq:3+1DomWall}
    \mathcal{L}_{2+1\;DW} = \bar{\phi}[i\gamma^{\hat{I}}\partial_{\hat{I}}-i\lambda a\gamma^{\hat{3}}-m]\phi,
\end{align}
where $I$ runs from $0,1,2.$
 
In this four-component representation of 2+1 spacetime dimensions, chirality remains well defined owing to the existence of a matrix that anticommutes with gamma matrices appearing in the kinetic operator of Eq. \eqref{eq:3+1DomWall} \cite{Anguiano2005, Pisarski}. One such matrix is $\gamma^{\hat{3}}$. Therefore, the ASDW Lagrangian again gives rise to a pseudoscalar mass $m_5 = \lambda a$. Having shown that this holds for both odd- and even-dimensional spacetimes, this provides evidence that a pseudoscalar mass is generically present in the effective theory for an ASDW.

We can still rewrite the effective Lagrangian Eq. \eqref{eq:3+1DomWall} with the canonical two-component spinors in 2+1 dimensions. Before proceeding, recall first that there are two inequivalent irreducible representations of the Clifford algebra in 2+1D spacetime. This is because, in two spatial dimensions, there are no transformations that relate solutions of different spins, so they are separated into inequivalent representations \cite{Anguiano2005, Applequist}. Furthermore, we expect that these two inequivalent representations are combined in the four-component representation used in Eq. \eqref{eq:3+1DomWall}, since in the 3+1-dimensional bulk there are both spin-up and spin-down solutions. 

To obtain the two two-component spinors of inequivalent representations as the upper and lower two components of the four-component spinor in Eq. \eqref{eq:3+1DomWall}, we must first change from the current chiral representation of the gamma matrices to a block-diagonal one, like
\begin{align}
	\tilde{\gamma}^{\hat{0}}=\begin{pmatrix}
		\sigma^3 & 0 \\
		0 & -\sigma^3
	\end{pmatrix}, \; \tilde{\gamma}^{\hat{1}}=\begin{pmatrix}
		i\sigma^1 & 0 \\
		0 & -i\sigma^1
	\end{pmatrix}, \; \tilde{\gamma}^{\hat{2}}=\begin{pmatrix}
		i\sigma^2 & 0 \\
		0 & -i\sigma^2
	\end{pmatrix}, \; \tilde{\gamma}^{\hat{3}}=\begin{pmatrix}
		0 & -1 \\
		1 & 0
	\end{pmatrix}.
\end{align}
This is necessary because the chiral representation mixes the two-component inequivalent representations, and expanding Eq. \eqref{eq:3+1DomWall} directly leads to a Lagrangian whose 2-dimensional block gamma matrices do not anticommute.

The transformation to this new representation $\tilde{\gamma}$ from the chiral representation $\gamma$ is given by the unitary transformation $\tilde{\gamma}=S^{-1}\gamma S$ where 
\begin{align}
	S = \frac{1}{2}\begin{pmatrix}
		(\sigma^1+\sigma^2) & i(-\sigma^1+\sigma^2) \\
		(\sigma^1+\sigma^2) & i(\sigma^1-\sigma^2)
	\end{pmatrix}
\end{align}
This is a unitary operator, and hence a symmetry of the Lagrangian. Performing the transformation $\psi\to S\psi=(\phi_A, \phi_B)^T$, we find the 2+1 ASDW Lagrangian written with two-component spinors
\begin{align}
\mathcal{L}_{2+1\;DW} = \bar{\phi}_A(i\gamma^j_A\partial_j -m)\phi_A + \bar{\phi}_B(i\gamma^j_B\partial_j  - m)\phi_B + i\lambda a(\bar{\phi}_A\phi_B + \bar{\phi}_B\phi_A).
\end{align}

As mentioned, we see that the effective Lagrangian contains two spinors, $\phi_A$ and $\phi_B$, belonging to the inequivalent irreducible representations
\begin{align}
`   \gamma^{\hat{j}}_{A/B} = \pm(\sigma^3, i\sigma^1, i\sigma^2),  
\end{align}
coupled via the pseudoscalar mass. 

As in the case of 1+1D ASDW, it is possible to get effective massless fermions in this 2+1 domain wall despite the fermions being massive in the bulk. From Eq. \eqref{eq:3+1DomWall}, with the chiral representation, we can have solutions in the domain wall obeying an effective  Weyl equation $i\gamma^{\hat{j}}\partial_{\hat{j}} \phi=0$ if
\begin{equation}\label{eq:chiral2+1ASDW}
	\gamma^{\hat{3}}\phi = \frac{im}{\lambda a} \phi.
\end{equation}
Given that $\gamma^{\hat{3}}$ has eigenvalues of $\pm i,$ the right-hand side of the preceding equation requires $\lambda a= m$. Plugging in $\lambda=a^{-1}\sqrt{m^2 + k_1^2 + k_2^2 -\omega^2}$ in this relation implies that there is a massless dispersion $\omega = \pm\sqrt{k_1^2 + k_2^2}.$ Furthermore, with $\gamma^{\hat{3}}$ effectively being a chiral operator, Eq. \eqref{eq:chiral2+1ASDW} dictates that any effective massless fermions localized on the domain wall will strictly be right-chiral; massless solutions with the opposite chirality do not exist. Similar to the previous discussion in 1+1 ASDW, massless fermions would be left-chiral if $a < 0.$

Finally, there are again Majorana fermions within these solutions for the case where $\omega = k_1 = k_2 = 0$, and the two components of $\chi$ are equal. For example, if $\chi = (1,1)^T$, the solutions Eqs. \eqref{eq:3+1genSolns} will have the form
\begin{align}
	&u_{-\lambda}(\omega=k_1=k_2=0) =\left(
		1,
		1,
		i,
		-i
	\right)^T \\
	&v_{\lambda}(\omega=k_1=k_2=0)
	= \left(
		-i,
		i,
		1,
		1
	\right).
\end{align}
Under the  conjugation operator
\begin{align}
    CK=
    \begin{pmatrix}
        0 & -i\sigma^y \\
	i\sigma^y & 0
    \end{pmatrix}K,
\end{align}
these spinors only change by a phase 
\begin{align}
	Cu_{-\lambda}^* = i u_{-\lambda},\;\;\;\;\;
	Cv_{\lambda}^* =  i v_{\lambda},
\end{align}
indicating that they are Majorana modes.

\section{Spherically-symmetric ASDW}
Finally, we look at the case of a spherically-symmetric ASDW, where the transition between antispacetime and spacetime happens radially. This is a toy model where there are packets or domains of antispacetimes within spacetime or vice-versa. Placing the ASDW at $r=r_0,$ we use the following vielbeins 
\begin{align}
    e_0^{\hat{0}}=1,\;\;  e_r^{\hat{r}} =\tanh[a(r-r_0)]\equiv f(r)^{-1},\;\; e_\theta^{\hat{\theta}} = r,\;\; e_\phi^{\hat{\phi}} = r\sin\theta.
\end{align}

Again, we are considering both signs of $a$: $a>0$ corresponds to an antispacetime region enclosed by spacetime, whereas $a<0$ corresponds to a spacetime region enclosed by antispacetime.

The nonvanishing spin connection coefficients in this background are 
    \begin{align}
    \omega_{2\hat{1}\hat{2}} = \coth[a(r-r_0)], \quad \omega_{3\hat{1}\hat{3}} = \coth(a(r-r_0))\sin\theta,\quad \omega_{2\hat{1}\hat{2}} = \cos\theta
    \end{align}
    and the spinorial affine connection or Fock-Ivanenko coefficients are
    \begin{equation}
    	\begin{aligned}
    		\Gamma_2 = \frac{1}{4}\coth(a(r-r_0))\qty[\gamma^{\hat{r}},\gamma^{\hat{\theta}}], \quad \Gamma_3 = \frac{1}{4}\coth(a(r-r_0))\sin\theta\qty[\gamma^{\hat{r}},\gamma^{\hat{\phi}}] + \frac{1}{4}\cos\theta\qty[\gamma^{\hat{\theta}},\gamma^{\hat{\phi}}].
    	\end{aligned}
    \end{equation}
    For the flat gamma matrices, we use the Dirac representation with the spatial part rotated into the spherical coordinate system
    \begin{equation}
    	\begin{aligned}
    		\gamma^{\hat{0}} = \begin{pmatrix}
    			1 & 0 \\
    			0 & -1
    		\end{pmatrix}, \quad \gamma^{\hat{r}} = \begin{pmatrix}
    			0 & \sigma^r \\
    			-\sigma^r & 0
    		\end{pmatrix},
    		\quad\gamma^{\hat{\theta}} = \begin{pmatrix}
    			0 & \sigma^\theta \\
    			-\sigma^\theta & 0
    		\end{pmatrix}, \quad \gamma^{\hat{\phi}} = \begin{pmatrix}
    			0 & \sigma^\phi \\
    			-\sigma^\phi & 0
    		\end{pmatrix}.
    	\end{aligned}
    \end{equation}
    Here we used the spherical representation of the Pauli matrices
    \begin{equation}
    	\begin{aligned}
    		&\sigma^r = \vec{\sigma}\cdot\hat{r} = \begin{pmatrix}
    			\cos\theta & e^{-i\phi}\sin\theta \\
    			e^{i\phi}\sin\theta & -\cos\theta
    		\end{pmatrix}, \\ 
    		&\sigma^{\theta} = \vec{\sigma}\cdot\hat{\theta} = \begin{pmatrix}
    			-\sin\theta & e^{-i\phi}\cos\theta \\
    			e^{i\phi}\cos\theta & \sin\theta
    		\end{pmatrix}, \\ 
    		&\sigma^{\phi} = \vec{\sigma}\cdot\hat{\phi} = \begin{pmatrix}
    			0  & -ie^{-i\phi} \\
    			ie^{i\phi} & 0
    		\end{pmatrix},
    	\end{aligned}
    \end{equation}
    where $\vec{\sigma}=(\sigma^1,\sigma^2,\sigma^3).$
    With these, the Dirac equation can be written as
    \begin{align}\label{eq:sphDiracEqn}
    	&\Bigg(i\gamma^{\hat{0}}\partial_t + i\coth[a(r-r_0)]\gamma^{\hat{r}}\qty(\partial_r + \frac{1}{r}) + \frac{i}{r}\gamma^{\hat{\theta}}\qty(\partial_\theta + \frac{\cot\theta}{2}) +\frac{i}{r\sin\theta}\gamma^{\hat{\phi}}\partial_\phi - m\Bigg)\Psi = 0.
    \end{align}
    We now exploit the spherical symmetry of the system to recast the Dirac equation into a differential equation involving only the radial coordinate. We begin by redefining the Dirac field as $\Psi=\left(r\sqrt{\sin\theta}\right)^{-1}\psi$
    and multiplying Eq. \eqref{eq:sphDiracEqn} on the left by $r\sqrt{\sin\theta}$ to get rid of the extra terms    \begin{align}\label{eq:sphDiracEqnNoTerms}
    	\Big(i\gamma^{\hat{0}}\partial_t + if(r)\gamma^{\hat{r}}\partial_r + \frac{i}{r}\gamma^{\hat{\theta}}\partial_\theta  +\frac{i}{r\sin\theta}\gamma^{\hat{\phi}}\partial_\phi - m\Big)\psi = 0.
    \end{align}
    
    Being spherically symmetric, this Dirac equation should have solutions that are eigenstates of parity, total angular momentum $\vb{J},$ and $z$-component of the angular momentum $J_z$ \cite{Sakurai}. This is satisfied by the spherical spinor functions $\mathscr{Y}^{j,m_j}_l(\theta,\phi)$, which will form the angular part of the solution. Here, $l, m_j,$ and $j$ are the orbital angular momentum, total magnetic, and total angular momentum quantum numbers, respectively. We write the solution as 
    \begin{align}\label{eq:sphericalAnsatz}
    	\psi = e^{-i\omega t}\qty(u(r) \mathscr{Y}^{j,m_j}_{l=j-\frac{1}{2}}, -iv(r) \mathscr{Y}^{j,m_j}_{l={j+\frac{1}{2}}})^T,
    \end{align}
    where $u(r)$ and $v(r)$ are the radial parts of the solution. Plugging this into Eq. \eqref{eq:sphDiracEqnNoTerms} and noting that the angular momentum operator is 
    \begin{align}
        \hat{L} = -i\left(\hat{\phi}\;\partial_\theta - \hat{\theta}\;\frac{1}{\sin\theta}\partial_\phi\right),
    \end{align}
    the Dirac equation becomes
    \begin{equation}
    	\begin{aligned}
    		(\omega-m)u(r)\mathscr{Y}^{j,m}_{l=j-\frac{1}{2}} + \sigma^r \qty(f(r)\partial_r-\frac{1}{r}\vec{\sigma}\cdot\hat{L})v(r)\mathscr{Y}^{j,m}_{l=j+\frac{1}{2}} = 0 \\
    		(\omega+m)v(r)\mathscr{Y}^{j,m}_{l=j+\frac{1}{2}} - \sigma^r \qty(f(r)\partial_r -\frac{1}{r}\vec{\sigma}\cdot\hat{L})u(r)\mathscr{Y}^{j,m}_{l=j-\frac{1}{2}} = 0.
    	\end{aligned}
    \end{equation}
    To isolate the remaining equation for the radial part, we use the following properties of the spinor spherical functions when acted on by the operators $\vec{\sigma}\cdot \hat{L}$ and $\sigma^r$ \cite{Sakurai}
    \begin{align}                     \vec{\sigma}\cdot\hat{L}\;\mathscr{Y}^{l,m}_j = \kappa(j,l) \; \mathscr{Y}^{l,m}_j.
    \end{align}
    Here $\kappa(j,l)$ is defined as
    \begin{align}
    	\kappa(j,l)=\begin{cases}
    		-(\zeta+1), \;\;\; &\text{for } l = j+\frac{1}{2} \\
    		\zeta-1, \;\;\; &\text{for } l = j-\frac{1}{2}
    	\end{cases},
    \end{align}
    where $\zeta = j+\frac{1}{2}.$ Meanwhile for $\sigma^r$ ,
    \begin{align}
    	\sigma^r\;\mathscr{Y}^{j,m_j}_{l=j\pm\frac{1}{2}} = - \mathscr{Y}^{j,m_j}_{l=j\mp \frac{1}{2}}.
    \end{align}
    With these properties, we obtain the coupled differential equations for the radial part of the solution     
    \begin{align}
    \label{eq:coupledDE1}
    	(\omega-m)u(r)-\left[f(r)\frac{d}{dr}+\frac{\zeta+1}{r}\right]v(r)=&0\\
    \label{eq:coupledDE2}
    	(\omega+m)v(r)+\left[f(r)\frac{d}{dr}-\frac{\zeta-1}{r}\right]u(r)=&0.
    \end{align}
    Note that instead of the ansatz Eq. \eqref{eq:sphericalAnsatz}, we could have swapped the spherical harmonic functions in the upper and lower components, 
    \begin{align}\label{eq:secondSphericalAnsatz}
    	\psi = e^{-i\omega t} \qty(u(r) \mathscr{Y}^{j,m_j}_{l=j+\frac{1}{2}}, -iv(r) \mathscr{Y}^{j,m_j}_{l={j-\frac{1}{2}}})^T.
    \end{align}
    This solution behaves differently under parity compared to the ansatz Eq. \eqref{eq:sphericalAnsatz}. In this case, the radial differential equation can be found by the replacement $\zeta\to-\zeta$ in Eqs. \eqref{eq:coupledDE1}-\eqref{eq:coupledDE2}.
   
    It is possible to obtain the permissible values of the parameters $a$ and $\zeta$ for a localized solution by looking at the asymptotic behavior of Eqs. \eqref{eq:coupledDE1} and \eqref{eq:coupledDE2} and imposing regularity and normalizability. 

    First, we consider the limit as $r\to0,$ where $f(r)\to -\coth(ar_0)=-\text{sgn}(a)\coth(|a|r_0)=-C.$ Note that here we defined $C$ such that its sign corresponds to the sign of $a$. Decoupling the radial differential equations in this asymptotic region gives
    \begin{align}
        &r^2C^2 u''-2rC u' - [C(\zeta-1)+(\zeta^2-1)]u=0, \label{eq:uDE0lim} \\
        &v(r) = -\frac{1}{\omega+m}\left(-Cu'-\frac{\zeta-1}{r}u\right), \label{eq:vDE0lim}
    \end{align}
    where the primes denote derivatives in $r$. Eq. \eqref{eq:uDE0lim} is a second-order ordinary differential equation, so we get two sets of solutions:
    \begin{align}\label{eq:solnSets1}
       u_1 \sim r^{\frac{\zeta+1}{C}+1}, \; v_1 \sim r^{\frac{\zeta+1}{C}}
    \end{align}
    and
    \begin{align}\label{eq:solnSets2}
       u_2 \sim r^{-\frac{\zeta-1}{C}}, \; v_2 \sim r^{1-\frac{\zeta-1}{C}}.
    \end{align}
    Since this is in the limit as $r\to0,$ we must impose restrictions on $C$ and $\zeta$ to have regular solutions. For $C>0,$ the first set is regular for any $\zeta,$ and the second set is regular only for $\zeta=1.$ On the other hand, for $C<0,$ only the second set of solution could be made regular without any restriction for $\zeta$. Note that with the replacement $\zeta\to-\zeta,$ we would still obtain the solutions Eqs. \eqref{eq:solnSets1}-\eqref{eq:solnSets1} and end up with the same conclusions.
    
    Meanwhile, in the asymptotic region where $r\to \infty,$ $\coth(a(r-r_0))\to 1$ and $\frac{1}{r}\to 0,$ and the decoupled radial differential equations become
    \begin{equation}
    	\begin{aligned}
    		u'' &= (m^2-\omega^2) u \\
    		v &= -(\omega +m)^{-1} u'.
    	\end{aligned}
    \end{equation}
    The solution behaves as
    \begin{equation}\label{eq:asymp2}
    	\begin{aligned}
    		&u \sim B_1 e^{-k r} + B_2e^{k r}\\
    		&v \sim k(\omega +m)^{-1}\qty(B_1 e^{-k r}-B_2e^{kr}),
    	\end{aligned}
    \end{equation}
    where $k = \sqrt{m^2-\omega^2}.$ We have bound solutions when $m^2 > \omega^2$ that decay exponentially as $r\to\infty$, after dropping the exponentially diverging term.  
    
    %We note that the solutions being regular at these points, along with what we show in the next section that the solution is regular at $r=r_0,$ we have shown that the solution is normalizable at these singular points. There will be a discrete spectrum for this because not for arbitrary $\omega$ can be regular at $r=0$ and also be deacaying as $r\to \infty.$
    
    We solve Eqs. \eqref{eq:coupledDE1} and \eqref{eq:coupledDE2} using series solution about the ASDW at $r_0$, which is a regular singular point as we show in Appendix \ref{sec:seriesSoln}. Figure \ref{fig:plots} (b) shows the radial probability density for $\zeta=1$ and different values of $a > 0$. To generate this plot, we first numerically solved for $u(r)$ in the domain $[\epsilon,r_0-\epsilon] \cup [r_0+\epsilon,r_0+5]$ using the second-order ODE. We treated this as an initial value problem, where function values and derivatives of the series solution at $r=r_0\pm\epsilon$ were used as the initial conditions. A piecewise function is then created with this numerical solution and the series solution, where the latter is used in the remaining domain $(r_0-\epsilon, r_0+\epsilon)$. For the plot, we used $\epsilon=10^{-6}$, and the series solution was truncated at order 12. 
    
    Figure \ref{fig:plots} (b) demonstrates that localization becomes more pronounced as $a$ increases. Recall that with Standard Model fermions and the assumption that the length scale $a^{-1}$ is dictated by quantum gravity effects and therefore resides at the quantum gravity scale,
    the relevant regime in our case is $a\gg m$. Therefore, we expect fermions to be extremely localized on the ASDW. Finally, notice that the right tail end of the blue curve does not vanish exponentially as $r\to \infty.$ This is because we used an arbitrary value for $\omega,$ which is not necessarily an eigenenergy of the differential equations. However, this plot is sufficient to qualitatively demonstrate the dependence of the localization on $a$. The presence of localized states is guaranteed, as discussed in Appendix \ref{sec:seriesSoln}.
    
    \section{ASDW Analogues in Superfluid He$^3$}
    \label{sec:ASDW Analogues}
    While ASDWs are, in principle, detectable via fermionic probes, their direct experimental observation remains challenging. It is remarkable that an analogue of antispacetime can be realized in superfluid $^3\mathrm{He}$ \cite{Salomaa1988CosmiclikeDomainWalls,Volovik1990,Jacobson1998EffectiveSpacetime,Silveri2014HardDomainWalls, Nissinen2018, Volovik2019,Volovik2021, Rudd2021StrongCouplingDomainWalls}. At low energies, the quasiparticle excitations in this system obey an effective Dirac equation in curved spacetime, with the orbital structure of the superfluid order parameter acting as the emergent vielbein field. These are experimentally observed \cite{Makinen2019}, allowing the investigation of fermion behavior in antispacetime analogues  under controlled laboratory conditions. 
    
    It is therefore important to discuss our results above in the context of this analogue antispacetimes. First, we contrast the ASDW in Eq. \eqref{eq:asdw1+1} with the realization in superfluid $^3$He, where the vielbein takes the form $e_{1}^{\hat{1}}=\coth(ax)$, instead of the vielbein $e_{1}^{\hat{1}}=\tanh(ax)$ considered in this work. The former yields the effective metric $ds^{2}=dt^{2}-\coth^{2}(ax),dx^{2}$ \cite{Volovik1999VierbeinWalls, Jacobson1998EffectiveSpacetime}. In this analogue system, the characteristic width of the wall is set by the superfluid coherence length \cite{Salomaa1988CosmiclikeDomainWalls,Silveri2014HardDomainWalls,Rudd2021StrongCouplingDomainWalls}, whereas for a genuine ASDW the scale $a^{-1}$ could be determined by a quantum-gravity length scale. In the superfluid analogue, the wall lies at an infinite proper distance from any point $x\neq0$, and the effective speed of \lq\lq light\rq\rq vanishes as the wall is approached. Consequently, unlike the ASDW considered here, this analogue wall is inaccessible to observers from either side \cite{Volovik1999VierbeinWalls}. 
    
    Furthermore, our reflectionless scattering that leads to particle-hole conversion can be compared to the case in the superfluid $^3$He antispacetime analogues. \cite{Volovik1999VierbeinWalls, Jacobson1998EffectiveSpacetime}. Here, the regime is typically $m/a\gg1$, whereas probing the ASDW with Standard Model fermions (e.g., electrons) corresponds to 
    $m/a\ll1$, since 
    $a$ is likely set by a quantum-gravity energy scale. 
    Classically, in superfluid $^3$He, the analogue antispacetime and spacetime regions cannot communicate with each other. However, once quantum corrections are taken into account, the fermions are likewise transmitted across the wall without reflection \cite{Volovik1999VierbeinWalls}.
    
    We note that an accessible ASDW considered in this work may nevertheless be engineered through a spatial variation of chirality. Chiral domains have been observed \cite{Ikegami2013Chiral,Kasai2018}, and the chirality of the $^3$He-A phase can be tuned via rotation \cite{Walmsley2012} or magnetic fields \cite{Sato2008}.
    
    Lastly, we have shown that a pseudoscalar mass can emerge in the effective theory of fermions localized at the ASDW. When coupled to gauge interactions, this gives rise to phenomena associated with the strong CP problem and renders the $\theta$ vacuum physically observable. The superfluid $^3$He-A is known to host analogues of ASDW and emergent gauge symmetry \cite{Volovik2009}. These, coupled with the emergent pseudoscalar mass in the ASDW shown in this work, suggest a promising platform for investigating the strong CP problem in controlled, table-top experiments. 
    
    \section{Conclusions} 
    We have uncovered a broad range of phenomena for fermions in the presence of an ASDW. Specifically, scattering across the domain wall induces particle–hole conversion, and the interface hosts localized fermionic modes. These localized solutions admit Majorana zero modes, chiral fermions, and emergent pseudoscalar mass. These signatures are, in principle, experimentally accessible, opening a possible observational avenue for detecting antispacetime regions in our universe.

% While ASDWs are, in principle, detectable via fermionic probes, their direct experimental observation remains challenging. In superfluid $^3$He, analogue realizations of antispacetime exist, though the corresponding ASDW analogues take the form $e^{\hat{1}}_1 \sim \coth(az)$, instead of $e^{\hat{1}}_1\sim\tanh(az)$ considered here. An ASDW may nevertheless be engineered through a spatial variation of chirality. Chiral domains have been observed \cite{Ikegami2013Chiral,Kasai2018}, and the chirality of the $^3$He-A phase can be tuned via rotation \cite{Walmsley2012} or magnetic fields \cite{Sato2008}, suggesting a viable route to tabletop probes of fermions at such domain walls.

%%%%%%%%%%%%%%%%%%%%%%%%%%%%%%%%%%%%%%%%%%%%%%%%%
%%%%%%%%%%%%%%%%% APPENDIX %%%%%%%%%%%%%%%%%%%%%%
%%%%%%%%%%%%%%%%%%%%%%%%%%%%%%%%%%%%%%%%%%%%%%%%%

\appendix
\section{Calculations for the General Solutions in the Planar ASDW}
    \subsection{1+1D General Solution}\label{sec:1+1detailedSoln}
    The coupled differential equation Eq. \eqref{diraccomponents} has the solution
    \begin{align}\label{eq:1+1Orig}
    	\tilde{\psi}_{\pm}(\omega,x) =& c_{1\pm} \begin{pmatrix}
    		-i J_{\pm} \sinh(\lambda\ln\cosh(ax)) \\
    		\cosh(\lambda\ln\cosh(ax))
    	\end{pmatrix} + c_{2\pm}\begin{pmatrix}
    		J_{\pm} \cosh(\lambda\ln\cosh(ax)) \\
    		i\sinh(\lambda\ln\cosh(ax))
    	\end{pmatrix},
    \end{align}
    where $J_{\pm} = \frac{\lambda a}{m\mp\omega}$ and $\lambda = \frac{\sqrt{m^2-\omega^2}}{a}$. We have a total of 4 spinors, 2 for each energy sector. Using the identities
    \begin{align}\label{eq:identities}
    	\cosh\qty(\lambda \ln \cosh(ax)) =& \frac{\cosh^\lambda(ax)+\cosh^{-\lambda}(ax)}{2}; \\
    	\sinh\qty(\lambda \ln \cosh(ax)) =& \frac{\cosh^\lambda(ax)-\cosh^{-\lambda}(ax)}{2},
    \end{align}
    we rewrite the solutions as
    \begin{align}\label{eq:intermediateSoln}
    	\tilde{\psi}_{\pm} &= A_{\pm} \cosh^{-\lambda}(ax)\begin{pmatrix}
    		iJ_{\pm} \\
    		1
    	\end{pmatrix} + B_{\pm}\cosh^\lambda(ax)\begin{pmatrix}
    		-iJ_{\pm} \\
    		1
    	\end{pmatrix},
    \end{align}
    where $A_{\pm}=\frac{c_{1\pm}-ic_{2\pm}}{2}$ and $B_{\pm}=\frac{c^*_{1\pm}+ic^*_{2\pm}}{2}$. We can write the solution more compactly by admitting negative values for $\lambda=\pm\frac{\sqrt{m^2-\omega^2}}{a}$. Renaming constants, we choose the form of the solutions to be
    \begin{align}
    	\tilde{\psi}_+ = c_{1,\lambda}\cosh^{\lambda}(ax)\begin{pmatrix}
    		-\frac{i\lambda a}{m-\omega} \\
    		1
    	\end{pmatrix}, \quad	\tilde{\psi}_- = c_{2,\lambda}\cosh^{-\lambda}(ax)\begin{pmatrix}
    		\frac{i\lambda a}{m+\omega} \\
    		1
    	\end{pmatrix}.
    \end{align}
    For clarity, it must be emphasized that each energy mode above is comprised of two spinors, as in Eqs. \eqref{eq:1+1Orig} and \eqref{eq:intermediateSoln} because $\lambda$ can now take both positive and negative signs. It follows that there are different constants for spinors associated with different signs of $\lambda$, meaning that there are four independent coefficients in total ($c_{1,|\lambda|},c_{1,-|\lambda|},c_{2,|\lambda|},c_{2,-|\lambda|} $). Next, we can make $c_{1,\lambda}$ absorb an additional factor of $\frac{m+\omega}{i\lambda a}$ so that spinors would have the same structure, where the non-unit component becomes $\frac{i\lambda a}{m+\omega}$. This leads to the solution Eq. \eqref{eq:gensoln}. 
    
    We choose the normalization constant $N = \sqrt{\frac{m}{\omega} (m+\omega)},$ such that the spinors satisfy the orthogonality relations 
    \begin{align}
    	\overline{u}_\lambda u_\lambda &=-\overline{v}_\lambda v_\lambda = 2m \\
    	\overline{u}_\lambda v_\lambda &= u_{-\lambda}^\dagger v_\lambda = 2i\frac{m}{\omega}\lambda a \\
    	\overline{u}_{-\lambda}v_\lambda &= u^{\dagger}_\lambda v_\lambda = 0,
    \end{align}
    which are relativistically invariant and valid in the limit $m\to0.$ 
    
    \subsection{2+1D General Solution}\label{sec:2+1detailedSoln}
    Because the 2+1D Dirac equation \eqref{eq:2+1dirac} has planar symmetry in $x$, we can write the solution as 
    \begin{align}
	\psi(\vb{x},t) = \int\frac{\dd{k_1}}{\sqrt{2\pi}} \; \qty(\tilde{\psi}_+ e^{-i\omega t+ik_1x}+\tilde{\psi}_- e^{i\omega t-ik_1x}).
    \end{align}
    Plugging this back in allows us to decouple Eq.\eqref{eq:2+1dirac} into
    \begin{align}\label{eq:2dCoupledDiffEq}
    	&(\pm \omega -m) \tilde{\psi}_{A\pm} + \;(\pm ik_1 - i \coth(ay)\partial_y)\tilde{\psi}_{B\pm} = 0\\
    	&(\mp \omega -m) \tilde{\psi}_{B\pm} + (\pm ik_1 + i \coth(ay)\partial_y)\tilde{\psi}_{A\pm} = 0,
    \end{align}
    which has the solution
    \begin{align}
    	\tilde{\psi}_{\pm} = \frac{c_{1\pm}}{m \mp\omega} \begin{pmatrix}
    		-i\lambda a \sinh(F(y)) \pm i k_1\cosh(F(y))\\
    		(m \mp\omega)\cosh(F(y))
    	\end{pmatrix} + 
    	\frac{c_{2\pm}}{m \mp\omega} \begin{pmatrix}
    		\mp k_1 \sinh(F(y)) +  \lambda a\cosh(F(y)) \\
    		(m \mp\omega)i\sinh(F(y)) 
    	\end{pmatrix}, 
    \end{align}
    where $F(y) = \lambda\ln\cosh(ay)$, and $\lambda = \frac{\sqrt{m^2+k_1^2-\omega^2}}{a}.$ As done in the 1+1 D case, this can be rewritten, using the identities Eq. \eqref{eq:identities}, into
    \begin{align}
    	\tilde{\psi}_{\pm} = A_{\pm}\cosh^\lambda(ay) \begin{pmatrix}
    		\frac{\pm i(k_1\mp\lambda a)}{m\mp\omega} \\
    		1
    	\end{pmatrix}  + B_{\pm}\cosh^{-\lambda}(ay)\begin{pmatrix}
    		\pm\frac{i(k_1\pm\lambda a)}{m\mp\omega} \\
    		1
    	\end{pmatrix},
    \end{align}
    where $A_{\pm} = \frac{c_{1\pm}+ic_{2\pm}}{2}$ and $B_{\pm}=\frac{c_{1\pm}-ic_{2\pm}}{2}$. If we redefine $\lambda$ to also include negative values, $\lambda = \pm\frac{\sqrt{m^2+k_1^2-\omega^2}}{a}$, the solution can be rewritten more compactly as
    \begin{align}
    	\tilde{\psi}_+= a_{\lambda}\cosh^\lambda(ay)\begin{pmatrix}
    		\frac{i(k_1-\lambda a)}{m-\omega} \\
    		1
    	\end{pmatrix}, \quad
    	\tilde{\psi}_-=b^*_{\lambda}\cosh^{-\lambda}(ay)\begin{pmatrix}
    		\frac{-i(k_1-\lambda a)}{m+\omega} \\
    		1
    	\end{pmatrix}.
    \end{align}
    Here, the coefficients are relabeled into $a_{\lambda}$ and $b^*_{\lambda},$ denoting that there are different constants for different signs of $\lambda$. We can make them absorb additional factors for the  normalization constant $N$ and for the spinor part of $\tilde{\psi}_+$ to have the same form as that of the positive energy spinor in 1+1D, where the upper component is $1$. 

    \subsection{3+1D General Solution}
    With our representation, the Dirac equation is
    \begin{align}
    	\label{eq:3+1diracOrig}
    	[i\gamma^{\hat{I}}\partial_{\hat{I}}+i\gamma^{\hat{3}}\coth(az)\partial_z-m]\psi=0,
    \end{align}
    where the repeated index $I$ is summed only from zero to two. This has translation symmetries in $t, x,$ and $y$, so the solution can be written as
    \begin{align}\label{eq:3+1fourierExpansion}
    	\psi(t,\mathbf{x})=\int \frac{\dd{k_1}\dd{k_2}}{2\pi}\qty(e^{-i\omega t+ik_Ix^I}\tilde{\psi}_+(z)+e^{i\omega t-ik_jx^j}\tilde{\psi}_-(z)),
    \end{align}
    where $\tilde{\psi}_+(z)$ ($\tilde{\psi}_+(z)$) is the positive (negative) energy mode. From the form of the solutions in the lower-dimensional cases we take the ansatz
    \begin{align}
    	\label{eq:3+1ansatz1}
    	\tilde{\psi}_{\pm}(z)=\cosh^{-\lambda}(az)\phi_{\pm},
    \end{align}
    where $\phi_{\pm}$ is z-independent spinor. Subsituting this into \eqref{eq:3+1diracOrig} we see that we only get a nontrivial spinor if\begin{align}
    	\label{det1}
    	\det (\omega\gamma^0-\gamma^jk_j-\lambda ai\gamma^3-m)=0,
    \end{align}
    setting a condition on $\lambda=\pm\frac{\sqrt{m^2+k_1^2+k_2^2-\omega ^2}}{a},$ which again is similar to our solutions in the previous cases, with the addition of $k_j$ terms for the directions longitudinal to the domain wall. The spinor parts are obtained by expanding Eq. \eqref{eq:3+1diracOrig},
    \begin{align}
    	\label{eq:3+1expanded1}
    	&(\pm\omega\mp k_j\sigma^j - i\lambda a\sigma^3)\phi_{B\pm} - m\phi_{A\pm} = 0 \\
    	\label{eq:3+1expanded2}
    	&(\pm\omega\pm k_j\sigma^j - i\lambda a\sigma^3)\phi_{A\pm} - m\phi_{B\pm} = 0
    \end{align}
    and writing $\phi_{A\pm}$ in terms of $\phi_{B\pm}$ or the other way around. We have written the spinors in our solutions for the lower dimensional cases in such a way that the upper component of $\tilde{\psi}_+$ and the lower component of $\tilde{\psi}_-$ have simpler expressions than the lower part of $\tilde{\psi}_+$ and the upper part of $\tilde{\psi}_-$, respectively. Keeping consistent with this convention, we use Eq. \eqref{eq:3+1expanded2} for $\phi_+$ and  Eq. \eqref{eq:3+1expanded1} for $\phi_-$. We label the independent 2-spinor $\chi$. The solution is
    \begin{align}\label{eq:3+1modeSolns}
    	&\tilde{\psi}_+ =\cosh^{\lambda}(az) \begin{pmatrix}
    		\chi \\
    		\frac{(\omega+k^j\sigma_j- i\lambda a \sigma^3) }{m}\chi
    	\end{pmatrix} = \cosh^{\lambda}(az) u_{\lambda}\\
    	&\tilde{\psi}_-= \cosh^{-\lambda}(az)\begin{pmatrix}
    		\frac{(-\omega+k^j\sigma_j- i\lambda a \sigma^3) }{m}\chi\\
    		\chi 		
    	\end{pmatrix} = \cosh^{-\lambda}(az) v_{\lambda}.
    \end{align}

\section{Inclusion of negative energy solutions in scattering across ASDW }\label{sec:negativeEnergyscattering}
 We now include the negative-energy modes in the scattering problem for completeness. As we will see, they lead to the same results of particle-hole conversion across ASDW presented in the main body of the article, where only positive-energy modes were considered.

To start, recall that the asymptotic limit $x\to\pm\infty$ of the Dirac equation, Eq. \eqref{eq:1+1Diraceq}, is
\begin{equation}\label{eq:asymptoticDirac}
	(i\sigma^z\partial_t\pm\sigma^y\partial_x-m)\psi = 0,
\end{equation}
which is Minkowski, giving us plane-wave solutions. We write them as
\begin{align}\label{eq:scatteringSolnForm}
	\psi^{\pm\infty} = e^{-i\omega t + ikx}\phi^{\pm\infty}_{+} + e^{i\omega t-ikx}\phi^{\pm\infty}_{-}.
\end{align}
Here, $\omega > 0,$ so that the first and second terms are positive and negative energy modes, respectively. The superscripts $\pm\infty$ denote the asymptotic regions in which the solution satisfies the Dirac equation. Plugging this back into Eq. \eqref{eq:asymptoticDirac} yields the condition $\omega^2 = m^2 + k^2,$ where right now $k$ can take positive and negative values, so that there are left- and right-moving positive and negative energy solutions in Eq. \eqref{eq:scatteringSolnForm}. The solutions are
\begin{equation}\label{eq:AsymptoticSolns}
	\begin{aligned}
		\phi^{\pm\infty}_+ = \bar{N}\left(1, \; \mp\frac{k}{\omega + m}\right)^T, \quad
		\phi^{\pm\infty}_-=\bar{N}\left(\mp\frac{k}{\omega + m}, \; 1\right)^T
	\end{aligned}
\end{equation}
where the normalization constant is $\bar{N} = \sqrt{\omega +m}$, so that the solutions satisfy the following normalization and orthogonal relations
\begin{align}
	\bar{\phi}_{\pm}\phi_{\pm} = {\pm}2m, \;\;\; \bar{\phi}_+\phi_- = 0.
\end{align}
We ommitted the superscripts $\pm\infty$ above, since these conditions hold for the solutions in both asymptotic regions. 

We now redefine $k$ to be greater than zero. We will be explictily adding signs to denote its previously negative values. With the asymptotic solutions, Eq. \eqref{eq:AsymptoticSolns}, we write the incident $\psi_I$ and reflected $\psi_R$ plane waves at $x\to-\infty$ as
\begin{equation}\label{eq:negInfsoln}
	\begin{aligned}
		\psi_I =& \bar{N}\Bigg(I_+e^{-i\omega t+ikx}\begin{pmatrix}
			1 \\
			\frac{k}{\omega + m}
		\end{pmatrix}+I_-e^{i\omega t-ikx}\begin{pmatrix}
			\frac{k}{\omega + m} \\
			1
		\end{pmatrix}\Bigg) \\
		\psi_R =& \bar{N}\Bigg(R_+e^{-i\omega t-ikx}\begin{pmatrix}
			1 \\
			-\frac{k}{\omega + m}
		\end{pmatrix}+R_-e^{i\omega t+ikx}\begin{pmatrix}
			-\frac{k}{\omega + m} \\
			1
		\end{pmatrix}\Bigg),
	\end{aligned}
\end{equation}
with $I_{\pm}$ being the nonzero incidence amplitudes and $R_{\pm}$ as the reflection amplitudes.

To match the $x\to-\infty$ asymptotic form of the general solution, Eq. \eqref{eq:gensoln}, to Eq. \eqref{eq:negInfsoln}, we first apply the plane-wave dispersion $\omega^2=k^2 +m^2$, so that $\lambda=\pm ika^{-1},$ giving 
\begin{align}\label{eq:genkSubs}
	\psi &= c_{1,\pm k}e^{-i\omega t}\cosh^{\pm ik/a}(ax)N\begin{pmatrix}
		1 \\
		\mp\frac{k}{m+\omega}
	\end{pmatrix}  \nonumber \\
	&+ c_{2,\pm k}e^{i\omega t}\cosh^{\mp ik/a}(ax)N\begin{pmatrix}
		\mp\frac{k}{m+\omega} \\
		1
	\end{pmatrix}.
\end{align}

In the limit $x\to-\infty,$ Eq. \eqref{eq:genkSubs} becomes
\begin{align}\label{eq:-infGen}
	\psi=N&\Bigg[\frac{c_{1,k} e^{-i\omega t-ikx}}{2^{ik/a}}\begin{pmatrix}
		1 \\
		-\frac{k}{\omega + m}
	\end{pmatrix} + \frac{c_{2,k} e^{i\omega t+ikx}}{2^{-ik/a}}\begin{pmatrix}
		-\frac{k}{\omega + m} \\
		1
	\end{pmatrix} \Bigg] \nonumber \\
	&+ \frac{c_{1,-k} e^{-i\omega t+ikx}}{2^{-ik/a}}\begin{pmatrix}
		1 \\
		\frac{k}{\omega + m}
	\end{pmatrix} + \frac{c_{2,-k} e^{i\omega t-ikx}}{2^{ik/a}}\begin{pmatrix}
		\frac{k}{\omega + m} \\
		1
	\end{pmatrix} \Bigg].
\end{align}
Continuity between the solutions Eqs. \eqref{eq:-infGen} and \eqref{eq:negInfsoln} requires
\begin{align}\label{eq:-infCond}
	&\frac{Nc_{1,-k}}{2^{-ik/a}} = I_+\bar{N}, \quad \frac{Nc_{2,-k}}{2^{ik/a}} = I_+\bar{N} \\
	&\frac{Nc_{1,k}}{2^{ik/a}} = R_+\bar{N}, \quad \frac{Nc_{2,k}}{2^{-ik/a}} = R_+\bar{N}.
\end{align}
Now considering the region $x\to\infty$, if we constructed the transmitted solution $\psi_T$ with only right-moving plane waves from Eq. \eqref{eq:AsymptoticSolns}, we have
\begin{align}
	\psi_T = \bar{N}\Bigg(T_{R+}e^{-i\omega t+ikx}\begin{pmatrix}
		1 \\
		-\frac{k}{\omega + m}
	\end{pmatrix}+T_{R-}e^{i\omega t-ikx}\begin{pmatrix}
		-\frac{k}{\omega + m} \\
		1
	\end{pmatrix}\Bigg),
\end{align}
where $T_{R\pm}$ are the transmission amplitudes for right-moving positive and negative-energy modes.

Meanwhile, in the limit as $x\to \infty$, the general solution Eq. \eqref{eq:genkSubs} becomes
\begin{align}\label{eq:+infGen}
	\psi=N&\Bigg[\frac{c_{1,k} e^{-i\omega t+ikx}}{2^{ik/a}}\begin{pmatrix}
		1 \\
		-\frac{k}{\omega + m}
	\end{pmatrix} + \frac{c_{2,k} e^{i\omega t-ikx}}{2^{-ik/a}}\begin{pmatrix}
		-\frac{k}{\omega + m} \\
		1
	\end{pmatrix} \Bigg] \nonumber \\
	&+  \frac{c_{1,-k} e^{-i\omega t-ikx}}{2^{-ik/a}}\begin{pmatrix}
		1 \\
		\frac{k}{\omega + m}
	\end{pmatrix} + \frac{c_{2,-k} e^{i\omega t+ikx}}{2^{ik/a}}\begin{pmatrix}
		\frac{k}{\omega + m} \\
		1
	\end{pmatrix} \Bigg].
\end{align}
If we imposed the continuity of these solutions at $x\to\infty$, we get the conditions 
\begin{gather}
	\frac{Nc_{1,k}}{2^{ik/a}} = T_+\bar{N}, \quad \frac{Nc_{2,k}}{2^{-ik/a}} = T_-\bar{N} \\
	c_{1,-k} = 0, \quad c_{2,-k}=0,
\end{gather}
which contradict Eqs. \eqref{eq:-infCond}, with our setup of $I_{\pm}\neq 0$. 

If instead we included left and right-moving plane-waves in the transmitted equation, with $T_{L\pm}$ denoting the transmission amplitudes for the left-moving positive and negative-energy modes, the asymptotic solution will be
\begin{align}
	\psi_T = &\bar{N}e^{-i\omega t}\qty[T_{R+} e^{ikx}\begin{pmatrix}
		1 \\
		-\frac{k}{\omega + m}
	\end{pmatrix} + T_{L+}e^{-ikx}\begin{pmatrix}
		1 \\
		\frac{k}{\omega + m}
	\end{pmatrix}] + \nonumber\\
	& \bar{N}e^{i\omega t}\qty[T_{R-} e^{ikx}\begin{pmatrix}
		-\frac{k}{\omega + m} \\
		1
	\end{pmatrix} + T_{L-}e^{-ikx}\begin{pmatrix}
		\frac{k}{\omega + m} \\
		1
	\end{pmatrix}].
\end{align}
The continuity of this new solution with Eq. \eqref{eq:+infGen}, gives the conditions
\begin{align}
	&T_{R+}\bar{N}=\frac{c_{1,k} N}{2^{ik/a}}, \quad T_{R-}\bar{N}=\frac{c_{2,k} N}{2^{-ik/a}}, \\
	&T_{L+}\bar{N} = \frac{c_{1,-k} N}{2^{-ik/a}}, \quad T_{L-}\bar{N} = \frac{c_{2,-k} N}{2^{ik/a}},
\end{align}
which do not run into contradictions as before. Combined with the conditions in Eq. \eqref{eq:-infCond}, these imply that
\begin{align}
	\begin{aligned}
		&T_{L+} = I_+, \quad T_L- = I_-,\\
		&R_+ = T_{R+}, \quad  R_- = T_{R-},
	\end{aligned}
\end{align}
where we see that this is nothing but the generalization of the results in the paper, that is,
a positive- (negative-) energy particle crossing the ASDW gets transmitted as a left-moving positive- (negative-) energy particle. And requiring the probability to be unity, forces us into a conclusion that there are no reflected and right-moving transmitted waves. Again, in keeping the result consistent with the causal strcuture of the metric, we interpreted such results as particle-hole conversion.

\section{Appendix for Spherically Symmetric ASDW}
        
    \subsection{Calculations in the behavior of solutions in the $r\to0$ limit}\label{sec:seriesSoln}
    Eq. \eqref{eq:uDE0lim} is Cauchy-Euler equation solved by $u_{1,2}=r^{\alpha_{1,2}}$, where the exponents are
    \begin{align}
        &\alpha_1= 1+ C^{-1}(\zeta+1),\\
        &\alpha_2 = -C^{-1}(\zeta-1).
    \end{align}
    We can use these solutions in Eq.\eqref{eq:vDE0lim} to get $v_1$
    \begin{align}
        v_1 \sim r^{\frac{\zeta+1}{C}}.
    \end{align}
    However, doing this for $v_2$ makes the RHS vanish, and we instead use the $r\to0$ limit of \eqref{eq:coupledDE1}
    \begin{align}
        -Cv' + \frac{\zeta + 1}{r}v = (\omega-m)u.
    \end{align}
    Plugging in $u_2$ here with the ansatz $v_2=Ar^{\beta},$ we get
    \begin{align}
        &A = \frac{m-\omega}{C-2\zeta}\\
        &\beta = 1-C^{-1}(\zeta-1).
    \end{align}
    
    For $C>0,$ the first set of solution Eq. \eqref{eq:solnSets1} will be regular for any value of $\zeta.$ Meanwhile, in the second set Eq. \eqref{eq:solnSets2}, $u_2$ will be regular for $\zeta\le1,$ while $v_2$ will be regular for $\zeta \le1+C.$ Since $\zeta=j+1/2\ge1,$ the only allowable value is $\zeta=1.$

    For $C<0,$ in the first set of solution, $v_1$ cannot be made regular for any allowable value of $\zeta.$ On the other hand, in the second set of solution $u_2$ will be regular for any allowable value of  $\zeta$, while $v_2$ will be regular $\zeta \ge1-|C|.$ Hence, any value of $\zeta$ is permissible.
    
    \subsection{Series solution for the linearized $e_r^{\hat{r}} = a(r-r_0)$ }\label{sec:seriesSoln}
    Eqs.\eqref{eq:coupledDE1}-\eqref{eq:coupledDE2} are analytically intractable, so we analyze the bound solutions near the interface $r_0$ by linearizing the radial vielbein $e_{r}^{\hat{r}} = \tanh(a(r-r_0)) \to a(r-r_0)$, such that $f(r)=\frac{1}{a(r-r_0)}.$ Decoupling the radial differential equations, we get a second-order differential equation in $u(r)$.
    \begin{equation}
    	\label{eq:uDE}
    	r^2 f^2(r) \dv[2]{u}{r} + f(r) \left(r^2 \dv{f(r)}{r} + 2r \right) \dv{u}{r} + \left(f(r)(\zeta - 1) + p^2 r^2 - \zeta^2 + 1 \right) u = 0,
    \end{equation}
    where $p^2 = \omega^2 - m^2,$ and an equation for $v\qty(r)$
    \begin{equation}
    	v(r) = -\frac{1}{\omega+m}\left(f(r)\dv{u(r)}{r}-\frac{\zeta-1}{r}u(r)\right).
    \end{equation}
    Observe that the following limits
    \begin{equation}
    	\begin{aligned}
    		&\lim_{r\to r_0} (r-r_0)\frac{\left(r^2 \dv{f(r)}{r} + 2r \right)}{r^2 f^2(r)} = -\frac{r_0^4}{a^3} \\
    		&\lim_{r\to r_0} (r-r_0)^2\frac{f(r)(\zeta - 1) + p^2 r^2 - \zeta^2 + 1}{r^2 f^2(r)} = 0
    	\end{aligned}
    \end{equation}
    are finite, meaning that $r_0$ is a regular singular point.
    
    We can, therefore, write $u(r)$ as a power series expansion about $r_0$
    \begin{align}
    	\label{eq:seriesSoln}
    	u = \sum_{n=0}^{\infty} b_n (r-r_0)^{q+n},
    \end{align}
    where $q$ is the indicial root. Multiplying Eq. \eqref{eq:uDE} by $a^3(r-r_0)^3,$ and rewriting its coefficients in powers of $(r-r_0)$, we get $P(r)\dv[2]{u}{r}+Q(r)\dv{u}{r} + R(r)u$ where the coefficients are
    \begin{align}\label{eq:DECoeffs}
    	P(r) &=a\qty((r-r_0)^3+2r_0(r-r_0)^2+r_0^2(r-r_0))  \\
    	Q(r) &= 2a^2(r-r_0)^3+(2a^2r_0-a)(r-r_0)^2-2ar_0(r-r_0)-ar_0^2 \\
    	R(r) &= p^2r^2a^3\qty((r-r_0)^5+2r_0(r-r_0)^4) + (r-r_0)^3(p^2a^3r_0^2-(\lambda^2-1)a^3) \nonumber \\
    	&\quad +(\lambda-1)a^2(r-r_0)^2.
    \end{align}
    
    Notice that $(q+n-1)$ will be the lowest power of $(r-r_0)$ from the $P(r)u''$ and $Q(r)u'$ terms given in Eq. \eqref{eq:DECoeffs}, while $(q+n+5)$ will be the highest power from $R(r)u$ term. We reindex the terms and write them as powers of $(r-r_0)^{q+n-1}$:
    \begin{equation}
    	\begin{aligned}
    		&(r - r_0)^{q+n-1} \Bigg[ 
    		\sum_{n=0}^{\infty} b_n r_0^2 a  (n+q)\left((n+q-1) - 1\right) + \sum_{n=1}^{\infty} 2 b_{n-1} r_0 a (n+q-1)\left((n+q-2) - 1\right) \\
    		&+ \sum_{n=2}^{\infty} b_{n-2}(n+q-2)a \big[\left(2 a r_0 - 1\right) +(n+q-3)\big] + \sum_{n=3}^{\infty} b_{n-3} \left[a^2(\lambda - 1) + 2 a^2(n+q-3)\right] \\
    		&+ \sum_{n=4}^{\infty} b_{n-4} a^3  \left(p^2 r_0^2 - (\lambda^2 - 1)\right) + \sum_{n=5}^{\infty} b_{n-5} 2 a^3 p^2 r_0  + \sum_{n=6}^{\infty} b_{n-6} a^3 p^2 \Bigg] = 0
    	\end{aligned}
    \end{equation}
    
    To solve the equation above, we group terms in the series with similar $b_n$'s ($n<6$), and demand that their coefficients vanish. This gives the indicial equation and coefficient values $b_1$ to $b_5$. For the indices $n\ge6,$ we group all the summations together and require it to vanish, which gives a recursion relation.
    
    From the $n=0$ part of the first term, we get two possible indicial values $q = 0, 2.$ Choosing a different indicial value leads to a different set of coefficient values and recursion relation. 
    For $q=0$:
    \begin{equation}
    	\begin{aligned}
    		b_1 = 0, \quad b_3& = -\frac{a b_0(\lambda-1)}{3r_0^2(2a-1)}, \quad b_4 = -\frac{a \left(a b_0 p^2 r_0^3-a b_0 \lambda ^2 r_0+a b_0 r_0-2 b_0 \lambda +4 b_2 r_0^2+2 b_0\right)}{8 r_0^3} \\ 
    		&b_5 = -\frac{a \left(2 a b_0 \lambda ^2 r_0-2 a b_0 \lambda  r_0+3 b_0 \lambda +b_2 \lambda  r_0^2-5 b_2 r_0^2-3 b_0\right)}{15 r_0^4}.
    	\end{aligned}
    \end{equation}
    Notice that $b_2$ is a free parameter. The equation that was supposed to give $b_2$ led to an identity, and hence, there is no restriction on the value of $b_2.$ The recursion relation that gives the coefficients $b_n$ is
    \begin{equation}
    	\begin{split}
    		b_n &= -\frac{1}{n(n-2)r_0^2} \bigg[ 
    		2(n-1)(n-3)r_0 b_{n-1} \quad + (n-2)(n-4+2ar_0) b_{n-2} 
    		+ a(2n-7+\lambda) b_{n-3} \\
    		&\quad - a^2(1-\lambda^2+p^2r_0^2) b_{n-4} 
    		+ 2a^2 p^2 r_0 b_{n-5} 
    		\bigg].
    	\end{split}
    \end{equation}
    Meanwhile, for $q=2$ the coefficient values are:
    \begin{equation}
    	\begin{aligned}
    		c_1=0, \quad c_2 =-\frac{ac_0}{2 r_0}, \quad c_3& = \frac{ac_0(5-\lambda)}{15r_0^2}, \quad c_4 = \frac{a c_0 \left(a r_0 \left(\lambda ^2-p^2 r_0^2+3\right)+2 (\lambda -3)\right)}{24 r_0^3} \\
    		&c_5 = -\frac{a c_0 \left(a \left(12 \lambda ^2-7 \lambda +35\right) r_0+6 (3 \lambda -7)\right)}{210 r_0^4},
    	\end{aligned}
    \end{equation}
    and the recursion relation is
    \begin{equation}
    	\begin{split}
    		c_n = -\frac{1}{n(n+2)r_0^2} \bigg[ 
    		&2(n^2-1)r_0 c_{n-1} + \left(n(n-2) + 2anr_0\right) c_{n-2} + a(2n-3+\lambda) b_{n-3} \\
    		&+ a^2(1-\lambda^2+p^2 r_0^2) c_{n-4}+ 2a^2 p^2 r_0 c_{n-5} 
    		\bigg].
    	\end{split}
    \end{equation}

    Overall, the solution as $r\to r_0$ is
    \begin{equation}
    \begin{aligned}\label{eq:asympr0}
        u(r) &= \sum_{n=0}^\infty b_n(r - r_0)^n 
	+ \sum_{n=0}^\infty c_n(r - r_0)^{2 + n} \\
		v(r) &= -(\omega + m)^{-1} \Bigg(\sum_{n=1}^\infty a^{-1} n b_n(r - r_0)^{n - 2} \nonumber
	\\& + \sum_{n=0}^\infty a^{-1}  c_n \left( 2 + n \right)(r - r_0)^n  +  -\left(\frac{\zeta-1}{r}\right)u(r) \Bigg)
    \end{aligned}
    \end{equation}

    From these results, we can argue about the existence of domain wall fermions parallel to the reasoning of \cite{Bagherian2024} for the presence of bound states in axion strings. We recapitulate their argument here. Eqs. \eqref{eq:coupledDE1}-\eqref{eq:coupledDE2} have regular singularities at $r=0$ and $r=r_0$ and an irregular singularity at infinity. A bound solution needs to be regular and continuous at these points, and at the same time vanish as $r\to\infty.$ We have shown that these are satisfied for specific values of $\zeta,$ $a$, and $k$. 

    Moreover, these differential equations are second-order and have only two linearly independent solutions, so there must be a mapping between the solutions as $r\to 0$ Eqs. \eqref{eq:solnSets1}-\eqref{eq:solnSets2}, solutions as $r\to r _0$ Eq. \eqref{eq:asympr0}, and solutions as $r\to\infty$ Eqs. \eqref{eq:asymp2} because they belong to the same solution space. There may be no bound states if solutions that are regular at $r=0$ and $r=r_0$ are the ones that diverge at infinity, or similarly, those that decay at infinity are not regular at $r=0$ and $r_0.$ In our case, at $r_0$ both of the linearly independent solutions are always regular and continuous. This means that the exponentially decaying solution at infinity will always be well-behaved as $r\to r_0.$ What could remain to be problematic is the solution behavior as $r\to 0.$
    
    For $C>0$ and $\zeta = 1,$ both of the linearly independent solutions at the origin are regular, and hence can be mapped to the decaying solution at infinity. On the other hand, for the cases of (1) $C>0$ with $\zeta \neq 1$, and (2) $C<0$ with $\zeta\ge1$ there is only one regular solution as $r\to 0.$ At best, bound solutions could only exist with a discrete spectrum of eigenenergies because not every $\omega$ can be regular at $r=0$ and $r=r_0$ and simultaneously decay at infinity.

\bibliographystyle{iopart-num}
\bibliography{apssamp}% Produces the bibliography via BibTeX.

\end{document}